\documentstyle[emulateapj]{article}

\slugcomment{To Appear in PASP, 2005 June issue}

\begin{document}

\title{Supernova 1954J (Variable 12) in NGC 2403
Unmasked\footnote{Based in part on observations with the NASA/ESA {\sl
Hubble Space Telescope}, obtained at the Space Telescope Science
Institute (STScI), which is operated by AURA, Inc., under NASA
contract NAS5-26555.}}

\author{Schuyler D.~Van Dyk}
\affil{Spitzer Science Center, Caltech, Mailcode 220-6, Pasadena CA  91125}
\authoremail{vandyk@ipac.caltech.edu}

\author{Alexei V.~Filippenko, Ryan Chornock, and Weidong Li}
\affil{Department of Astronomy, 601 Campbell Hall, University of
California, Berkeley, CA  94720-3411}
\authoremail{alex@astro.berkeley.edu, chornock@astro.berkeley.edu,
weidong@astro.berkeley.edu}

\and

\author{Peter M. Challis}
\affil{Harvard-Smithsonian Center for Astrophysics, 60 Garden St.,
Cambridge, MA 02138}
\authoremail{pchallis@cfa.harvard.edu}

\begin{abstract}

We have confirmed that the precursor star of the unusual Supernova
1954J (also known as Variable 12) in NGC 2403 survived what appears to
have been a super-outburst, similar to the 1843 Great Eruption of
$\eta$ Carinae in the Galaxy.  The apparent survivor has changed
little in brightness and color over the last eight years, and a Keck
spectrum reveals characteristics broadly similar to those of $\eta$
Car.  This is further suggested by our identification of the actual
outburst-surviving star in high-resolution images obtained with the
Advanced Camera for Surveys on the {\sl Hubble Space Telescope}.  We
reveal this ``supernova impostor'' as a highly luminous
(${M_V}^0\approx -8.0$ mag), very massive ($M_{\rm initial} \gtrsim
25\ M_{\odot}$) eruptive star, now surrounded by a dusty ($A_V \approx
4$ mag) nebula, similar to $\eta$ Car's famous Homunculus.

\end{abstract}


\section{Introduction}

The fate of the most massive stars in galaxies is not well understood.
Such stars have been theoretically linked to black-hole formation, to
some gamma-ray burst sources, and to early metal enrichment in the
Universe (e.g., Umeda \& Nomoto 2003).  In principle, stars with
initial masses $M \gtrsim 20$--$30\ M_{\sun}$ should pass through the
red supergiant phase, or directly to the short-lived luminous blue
variable (LBV) phase, on to the Wolf-Rayet (WR) phase (e.g., Langer et
al.~1994; Stothers \& Chin 1996), before exploding as supernovae (SNe;
e.g., Woosley, Langer, \& Weaver 1993) or ending as ``collapsars''
(e.g., MacFadyen \& Woosley 1999).  The Great Eruption of 1843 for
$\eta$ Carinae (see Davidson \& Humphreys 1997), where the star's
bolometric luminosity increased by a very large amount, demonstrated that 
some very massive stars go through spectacular eruptive phases of pre-SN 
mass ejection.  This instability may play a role in shaping the appearance
of the upper Hertzsprung-Russell (H-R) diagram (Humphreys \& Davidson
1994).  However, LBVs, and particularly phenomenal cases such as
$\eta$ Car, are extremely rare (Humphreys \& Davidson 1994); isolating
additional examples is therefore necessary for a better understanding of
the evolution of very massive stars.

The identification of a number of extragalactic objects as genuine SNe
(strictly defined to be the violent destruction of a star at the end of its
life) has been questioned by several teams of investigators (e.g., Goodrich et
al.~1989; Filippenko et al.~1995; Van Dyk et al.~2000, 2004; Smith, Humphreys,
\& Gehrz 2001; Wagner et al.~2004).  These authors have argued that the ``SN
impostors'' are more likely the pre-SN cataclysmic event of a very massive
star, analogous to $\eta$ Car (hence, the term ``$\eta$ Car analogs" or
``$\eta$ Car variables''; Humphreys, Davidson, \& Smith 1999).  Similarly to
$\eta$ Car, the precursor star is expected to survive the eruptive
super-outburst.  The prototypical examples are SN 1961V in NGC 1058 (Zwicky
1964; Goodrich et al. 1989; Filippenko et al. 1995; Humphreys et al. 1999; Van
Dyk, Filippenko, \& Li 2002) and SN 1954J in NGC 2403 (Humphreys et al. 1999;
Smith et al. 2001).  Five more-recent examples are SN 1997bs in M66 (Van Dyk et
al.~2000), SN 1999bw in NGC 3198 (Filippenko, Li, \& Modjaz 1999), SN 2000ch in
NGC 3432 (Filippenko 2000; Wagner et al. 2004), SN 2001ac in NGC 3504 (Matheson
et al. 2001), and SN 2002kg, also in NGC 2403 (Schwartz \& Li 2002).  The star
V1 in NGC 2366 (Drissen, Roy, \& Robert 1997; Drissen et al. 2001) may be an
additional member of this group.  For SN 1997bs (Van Dyk et al. 1999) and SN
2002kg (Weis \& Bomans 2005; Van Dyk et al. 2005), luminous, very massive
precursor stars have been directly identified.

Whereas the early-time optical spectrum of SN 1961V was dominated by
narrow emission lines of H, He~I, and Fe~II (indicating a relatively
low maximum expansion velocity of $\sim 2000$ km s$^{-1}$; Zwicky
1964), similar to what is seen for the ``Type IIn" SN classification
(Schlegel 1990; Filippenko 1997), no spectra of SN 1954J exist at a
comparable epoch. The identification of this latter event as a SN,
therefore, has always been debatable.  Kowal et al.~(1972) considered
it a SN, possibly of ``Type V" (Zwicky 1965), based on the relative
similarity of its light curve to the highly peculiar one of SN 1961V.
Tammann \& Sandage (1968) had identified the object as an irregular
blue variable, Variable 12 (V12; see their Figure 11 for the unusual
light curve).  In quiescence the star showed only minor variability,
with $B \approx 21$ mag, but subsequently in outburst it fluctuated
drastically in brightness over a 5-yr span, beginning in 1949,
reaching a maximum $B \lesssim 16.5$ mag in mid-to-late 1954, when
it was the brightest star in NGC 2403.  By late 1954, it had declined steeply
to a level fainter than its pre-outburst state.  (SN 1954J is still seen
brightly in the POSS-I plates, taken on 1955 January 29 UT!)
Humphreys et al.~(1999) considered this behavior for SN 1954J/V12 quite unlike
what is seen for an actual SN.

No survivor of an extragalactic super-outburst has yet been
unambiguously recovered years to decades afterwards.  The most
contentious case is SN 1961V (see Chu et al. 2004; Van Dyk
2005, and references therein), where an $\eta$ Car-like star has
been detected in the SN environment, but it is still uncertain if it
can be directly linked to a putative survivor.  A more clear-cut case
may be SN 1954J/V12, where Smith et al.~(2001) first identified,
through ground-based optical ($UBVRI$) and near-infrared ($JHK$)
imaging, an object they considered the post-outburst survivor star.

Here we contribute further imaging of SN 1954J/V12 from the ground
with the Keck-I 10-m telescope and the 2.56-m Nordic Optical Telescope,
confirming the source identified by Smith et al.~(2001).
More importantly, we demonstrate with the superior spatial resolution
of the {\sl Hubble Space Telescope\/} ({\sl HST}) that the object seen
from the ground is actually a group of stars.  Furthermore, we are
able to isolate a star within the group as the likely candidate for
the 1954 super-outburst survivor.  By combining these data with an optical spectrum
from the Keck-II telescope, we argue that this star is a rare, very
luminous and massive, $\eta$ Car-like variable star.

We assume throughout this paper that the true distance modulus to NGC
2403 is $\mu_0=27.48$ mag (Freedman et al. 2001).  At this distance,
1\arcsec\ corresponds to 15.2 pc.

\section{Observations and Analysis}

\subsection{Keck Imaging}

SN 1954J/V12 was imaged in $VRI$ with the Low Resolution Imaging
Spectrometer (LRIS; Oke et al.~1995) on the Keck-I telescope on 1996
October 20 (UT dates are used throughout this paper).
Exposure times were 240 s in each band, and the seeing
was quite good in the $R$-band ($0{\farcs}63$), less so at $I$
($0{\farcs}74$), and mediocre at $V$ ($\sim 1{\farcs}3$, largely due to
less-than-optimal focus).  Standard image-calibration
procedures (bias subtraction, flat-fielding, etc.) were employed in
the reduction of these data.  We extracted photometry from these
images using IRAF\footnote{IRAF (Image
Reduction and Analysis Facility) is distributed by the National
Optical Astronomy Observatories, which are operated by the Association
of Universities for Research in Astronomy, Inc., under cooperative
agreement with the National Science Foundation.}/DAOPHOT (Stetson
1987) and fitting an
empirically derived model point-spread function (PSF) for each band.
Unfortunately, no photometric calibration was obtained during this night;
thus, we had to tie the photometry to the calibrated photometry obtained from
the Nordic Optical Telescope (NOT) images (see \S 2.2).

From the digitized red POSS-I image (the Digitized Sky Survey; DSS) we
measure an accurate absolute position for SN 1954J/V12 of
$\alpha$(J2000) = $7^h\ 36^m\ 55{\fs}36$, $\delta$(J2000) =
$+65\arcdeg\ 37\arcmin\ 52{\farcs}1$, with a total uncertainty of $\pm
0{\farcs}3$, using the Two Micron All Sky Survey (2MASS) as the
astrometric reference (see Van Dyk, Li, \& Filippenko 2003).  Applying
the same astrometric grid to the Keck images, we confirm in these
images at this position the source identified by Smith et al.~(2001;
see their Figure 2).

In Figure 1 (left panel)
we show the $R$-band image, revealing that the source appears extended
with an ``appendage'' $\sim 0{\farcs}7$ to the southwest.  The
appearance of the source changes in the $I$ band; see Figure 1 (right
panel).  Two closely spaced sources are identified by DAOPHOT in this
band, with the ``appendage'' source becoming more obvious, and the
main source to the north containing less light in this band.  We
identify the main source, seen clearly in the left panel, as the Smith
et al. (2001) object and list its photometry from the Keck images in Table 1,
along with the photometry from Smith et al.~(2001).

\subsection{Nordic Optical Telescope Images}

Larsen \& Richtler (1999) obtained $UBVRI$ (plus H$\alpha$) images of
NGC 2403 with the 2.56-m NOT on 1997
October 13, under very good
seeing conditions of $0{\farcs}8$.  These images were contributed to
the NASA/IPAC Extragalactic Database (NED) and posted for public
distribution on their website\footnote{http://nedwww.ipac.caltech.edu}.
We extracted photometry from the broad-band images,
first with a 4\arcsec\ aperture and then
using PSF fitting within IRAF/DAOPHOT.  The resulting
instrumental PSF-fitting magnitudes, after aperture corrections, were
calibrated using photometry of a number of isolated stars in the
images through a 20\arcsec\ aperture, which matches the aperture used
to establish the calibration through observations of standard stars by
Larsen (1999; his Table 1 and \S 3.1; we also first applied the
appropriate extinction corrections, supplied via a private
communication from S.~S.~Larsen).  We have compared the overall
photometry with the $BV$ bright-star photometry in Sandage (1984) and
$UBVR$ photometry in Zickgraf \& Humphreys (1991) and find very good
agreement across all bands.

The same astrometric grid applied to the DSS
and Keck images was also applied to the NOT images, again
confirming the Smith et al.~(2001) source.  The varying shape of the source as a
function of wavelength, as seen in the Keck images, is also seen in the NOT data.
In particular, the ``appendage'' is also detected in the NOT $I$-band image.
The $UBVRI$ photometry for the source from the
NOT images is given in Table 1.

\subsection{Keck Spectroscopy}

An optical spectrum of the SN 1954J/V12 site was obtained with the Echellette
Spectrograph and Imager (ESI; Sheinis et al.~2002) on the Keck-II telescope on
2002 November 9.56 (exposure time 2400~s).  The 1\arcsec\ slit included both
the Smith et al.~(2001) source and the ``appendage,'' but the seeing was poor
($\sim1.3''$) and the two sources were blended. Variable seeing (it was $\sim
0.8''$ during the observation of the standard star) and the presence of thin
cirrus compromised the absolute flux calibration to some degree.

Conventional data reductions were performed within IRAF, including bias
subtraction, flat-fielding, and wavelength calibration, while our own
IDL\footnote{IDL is the Interactive Data Language, a product of Research Systems, Inc.}
routines were used for flux calibration and removal of telluric lines (e.g.,
Matheson et al. 2000).  The extraction from the relevant order containing
H$\alpha$ (the only feature detected in the overall spectrum) is shown in
Figure 2.  Although the spectrum is somewhat noisy, broad-line emission with 
full-width at half-maximum (FWHM) of $\sim$1650 km s$^{-1}$ is clearly seen.  
The Gaussian velocity width, $\sigma$, of the line is then 
$\sim$700 km s$^{-1}$.  The equivalent width of this emission line is
$\sim$240~\AA.  We suspect the origin of this emission is either within the
extended source or from the appendage, but cannot tell from the spectrum alone
(we discuss this in more detail below).

\subsection{{\sl HST\/} Imaging}

The recent, bright SN 2004dj in NGC 2403 was imaged by program
GO-10182 (PI: Filippenko) with the Advanced Camera for Surveys (ACS)
on-board {\sl HST\/} on 2004 August 17.  Both the Wide Field Channel
(WFC) and the High Resolution Channel (HRC) were used, and the
resulting data immediately became publicly available in the {\sl
HST\/} archive.  Images were obtained with the WFC in the bands F475W,
F606W, F814W, and F658N, with respective exposure times of 600, 350,
350, and 650 s.  The imaging strategy for SN 2004dj was purposely
designed so that the site of SN 1954J/V12 would also be contained in
the WFC images.  We recombined the individual flat-fielded ``flt'' exposures
in all four bands at the Space Telescope Science Institute (STScI), in order to
more reliably remove cosmic-ray hit features from the images, and, in
effect, produce our own drizzled (Fruchter \& Hook 2002) ``drz'' images
independent of the {\sl HST\/}
data pipeline.  In Figure 3 we show the F606W, F814W, and F658N
images.  Comparison with Figure 1 (in particular, see the inset to
Figure 3a) clearly shows that the source identified from the ground is
actually a resolved group (or cluster) of stars.  We label in Figure 3
the four brightest members of the cluster, in decreasing order of
their F606W apparent brightness.

Since the stellar environment is crowded,
we measured the brightnesses of the four stars in the three broad-band
images and the one narrow-band image
using DAOPHOT, fitting the stellar profiles with PSFs, based on a
$0{\farcs}5$ aperture and generated
from the $\sim 10$ brightest, uncrowded stars in a $450 \times 300$
pixel area around the SN site\footnote{The location of the SN
site is so close to the edge of the WFC images, that the geometric
distortion added to artificial PSFs generated via TinyTim v6.2 (Krist
\& Hook 2003), appropriate for ACS, was too extreme to be useful for
the photometry.}.   We verified that this produces, to within the measurement
uncertainties, equivalent results
as measuring, first, the instrumental magnitudes through a $0{\farcs}15$
aperture and, second, correcting these magnitudes to those through a $0{\farcs}5$
aperture, at least for the two blue broad bands (aperture corrections for
the F814W band are more complex, due to the possible presence of the
red halo feature around stars; see Sirianni et al.~2005).  We determined that any
correction to the magnitudes for the degradation of
charge-transfer efficiency (CTE) is $<$0.6\% in the
three broad bands (see Riess \& Mack 2004).  The CTE correction for the F658N
band measurements, however, was more appreciable, at $\sim$2\%.

For the F475W, F606W, and F658N
measurements we next applied the correction from a $0{\farcs}5$ aperture to
infinite aperture from Table 5 in Sirianni et al.~(2005); the same is also true
for the blue Star 3 in the F814W band.  For Stars 1, 2, and 4 in the F814W
band we applied the corrections from Figure 12 in Sirianni et al.~(2005).
We then applied the zeropoints in Table 11 of Sirianni et
al.~to the corrected instrumental magnitudes, VEGAmag for the broad
bands and STmag for the narrow band.
The resulting flight system magnitudes
are given in Table 2.  We also transform the broad-band
magnitudes into the standard Johnson-Cousins system, following the translation
established by Sirianni et al., with the understanding that this transformation
is fraught with uncertainties, especially for objects with non-standard photospheres.
Furthermore, we do not know {\it a priori\/} the extinction toward the four stars, so
we perform the transformation for each star first, then establish the extinction,
contrary to the recommendation by Sirianni et al.
In Table 2 we list the resulting $BVI$ magnitudes for the four stars.

\section{Discussion}

We suggest that what is seen from the ground for SN 1954J/V12 is the
composite of the
stars resolved in the {\sl HST\/} images,
with the various individual stars contributing different amounts to
the overall brightness at different wavelengths.  This is borne out in
Table 1, where at $BVI$ we sum up the flux contributions from Stars 1,
3, and 4.  (When the WFC images are smoothed using a
Gaussian with widths corresponding to the ground-based seeing for the
Keck and NOT images, it is readily apparent that the Smith et al.~2001
source consists of these three stars.)  Star 2 is almost certainly the
``appendage'' to the main
source, as seen from the ground, and we therefore do not include the
contribution from its light.   These sums at each band agree quite
well with the brightness of SN 1954J/V12 in the ground-based
measurements.  However, we note that the $U$ and $I$ magnitudes from
Smith et al.~(2001) are generally fainter than the other measurements.
(Smith et al.~employed a 1\arcsec\ aperture with 2\arcsec\ seeing for
their measurements, and for the $I$ band the measurement is tied to one
comparison star, without an adequate set of observed standard stars,
which could account for some of the discrepancies.)
Also, our $B$ measurement from the 1997 image data and our $V$
measurement from the 1996 data are somewhat brighter than the ensemble
in that band, but we have already noted the degraded focus in
the Keck $V$-band image.  We cannot exclude that one or more of the
stars in the environment, even SN 1954J/V12 itself, is somewhat
variable.  However, it appears that the brightness and color of the
Smith et al. (2001) source have been relatively constant overall
between 1996 and 2004.

The properties (line width, equivalent width) of the broad H$\alpha$
emission line in SN 1954J/V12, as seen in the ground-based spectrum,
are strikingly similar to those of $\eta$ Car itself.  In Figure 4 we
directly compare the SN 1954J/V12 spectrum with a recent (1998 January 1)
spectrum of $\eta$ Car obtained with {\sl HST}/STIS by
the {\sl HST\/} Treasury Program (GO-9420; PI: K.~Davidson) on $\eta$
Car\footnote{The spectrum of ``Eta Car A'' was obtained from that Program's public
archive at http://etacar.umn.edu.}.  The broad components of the two
spectra agree quite well, although the narrower component, as well as
the pronounced P-Cygni profile in the broader component, seen in the
$\eta$ Car spectrum are both noticeably absent in the SN 1954J/V12
spectrum.

We contend that this late-time spectrum of SN 1954J/V12 is not of a
decades-old SN: The spectrum of SN 1957D in M83, of comparable age, is
dominated by lines due to oxygen with velocities in excess of
$\sim$2000 km s$^{-1}$ (Long, Winkler, \& Blair 1992).  We also
consider the probability of such a rare $\eta$ Car-like object being
in the immediate vicinity of SN 1954J/V12 (which photometrically
resembles $\eta$ Car; Humphreys et al.~1999), yet {\it not\/} being the SN
itself, to be extremely small.  It is most likely that one of the four
brightest stars at the SN site seen in the {\sl HST\/} images is the
source of the broad H$\alpha$ emission in the Keck spectrum, and that
it is the survivor of the SN 1954J/V12 outburst.  But, which one?

Examining Figure 3c we see that, within the environment, one star has
a clear, luminous excess of emission in the F658N band, relative to
the other three stars.  Comparing with Figures 3a and 3b, Star 4 is
exactly at the position of this emission source.  From Table 2 we see
that the STmag for Star 4 at F658N is $21.00 \pm 0.05$ mag (without any
continuum subtraction), which corresponds to an
equivalent flux density per unit wavelength of $1.45 \times 10^{-17}$
erg s$^{-1}$ cm$^{-2}$ \AA$^{-1}$.  We note that this agrees quite well
with the value of $f_\lambda$ near 6580~\AA\ in the spectrum shown
in Figure 2.  (This agreement assumes that the line flux has not changed
between 2002 November and 2004 August, and also may be fortuitous,
given the variable seeing conditions and thin cirrus present at the time of
the Keck observations.) Thus, we find it suggestive that the source of
the broad H$\alpha$ emission is indeed Star 4.

This is further supported by inspection of the 2-dimensional ESI spectrum of
the specific order (\#9) containing H$\alpha$.  In Figure 1a we label a star (``A''), to
the northeast of the SN environment, which also fell within the ESI slit.
Using this star as a fiducial, and measuring the spatial offset between the
spatial peak of this fiducial star's spectrum and that of the H$\alpha$ source,
we find $6{\farcs}7 \pm 0{\farcs}3$.  Measuring the offsets between this star
and stars in the grouping on the ACS/WFC images, we find $7{\farcs}2$ for Star
1, $7{\farcs}8$ for Star 2, $6{\farcs}8$ for Star 3, and $7{\farcs}0$ for Star
4 (with uncertainties of $<1$ WFC pixel, i.e., $\la 0{\farcs}05$).  We can
reject Stars 1 and 2 as the H$\alpha$ source, given their large displacement
from the fiducial.  Star 3, at face value, is in closest agreement with the
spectrum offset; however, we see very weak emission in the F658N image at its
position (see Table 2).  This, again, leaves Star 4 as the most likely
candidate for the emission seen in the spectrum, whose displacement from the
fiducial star is within the uncertainties in the measured offset.

What are the stars in the environment of SN 1954J/V12?  In particular,
are the characteristics of Star 4 consistent with an $\eta$ Car-like
star and the survivor of SN 1954J/V12, as suggested by the spectrum?
To provide possible answers to these questions, we place the four
stars (from the magnitudes in Table 2) on a [($B-V$), ($V-I$)] color-color
diagram, shown in Figure 5.  Also illustrated are the intrinsic color
tracks for dwarfs and supergiants from Drilling \& Landolt (2000), as
well as the reddening vector for the Cardelli, Clayton, \& Mathis
(1989) reddening law.  We note that, for these colors, the reddening
vector runs parallel to the stellar tracks.  At their exposure levels,
we would expect the ACS images to be sensitive generally to
supergiants and the bright end of the main sequence.

Stars 1 and 2 could be blue
supergiants experiencing local extinction $A_V \approx 7$ mag.
However, their absolute magnitudes, corrected for this extinction and
the true distance modulus, would then be an implausible ${M_V}^0
\approx -12$.  These two stars could possibly be reddened, luminous
yellow (G-type) supergiants; in this case, the required extinction
would be $A_V \approx 3$ mag, with ${M_V}^0 \approx -7.8$, $(B-V)_0
\approx 0.7$, and $(V-I)_0\approx 1.1$ mag.  However, such yellow
supergiants, or so-called hypergiants, are relatively rare (it seems 
even less likely that {\it both\/} stars would be hypergiants).
Alternatively, both Stars 1 and 2 have colors consistent with 
early-M supergiants (at least for $V-I$; the transformation to $B-V$
appears to be too red, but the uncertainty in this color for both
stars is large and still consistent with the supergiant track),
assuming
low reddening (the Galactic contribution to extinction toward
NGC 2403 is only $A_V = 0.13$ mag; Schlegel, Finkbeiner, \& Davis
1998), and we consider this the more likely case.

Star 3 could be a
blue star with $A_V \approx 1.4$ mag; its reddening-free colors would
be $(B-V)_0\approx -0.27$ and $(V-I)_0\approx -0.47$ mag; its absolute
magnitude, ${M_V}^0 \approx -6.0$, would imply that it is a hot
main-sequence dwarf.  However, Star 3 also has
colors consistent with early-A to early-F type with little reddening
(although the star's locus is offset in color by $\sim 1\sigma$ from
the stellar tracks in Figure 5).  Additionally, this star is $\sim$0.7
mag fainter in F658N than in F606W, implying a relatively high
H$\alpha$ absorption equivalent width, consistent with an A-type
star; using the STSDAS package SYNPHOT, together with
synthetic spectra from the Bruzual-Persson-Gunn-Stryker (BPGS)
Spectral Atlas\footnote{See the SYNPHOT
User's Guide for a description of the BPGS atlas;
http://www.stsci.edu/resources/software{\textunderscore}hardware/stsdas/synphot.},
we find that the flight-system colors for this 
star are most consistent with those of an early-to-mid A star.
Its inferred absolute magnitude is then only ${M_V}^0 \approx -4.7$.

With little reddening, Star 4 has colors generally consistent with
spectral type G (although its locus in Figure 5 is also offset in
color by $\sim 1\sigma$ from the stellar tracks).  Based on its
observed brightness, it would have to be a supergiant.  However, we
would still need to account for the presence of the broad H$\alpha$
line from a G-type supergiant.  The yellow supergiants
(hypergiants), such as $\rho$ Cassiopeiae (e.g., Lobel et al. 2003), do
show H$\alpha$ emission associated with a massive wind.  However, the
emission profile for $\rho$ Cas is quite different in shape, velocity
width, and equivalent width from what we see in the Keck spectrum for
Star 4.  Additionally, $\rho$ Cas, spectroscopically of late-F or
early-G type, has $A_V \approx 1$--2 mag (e.g., Joshi \& Rautela 1978)
and ${M_V}^0 \approx -7.8$ mag (e.g., Lobel et al.~2003), whereas Star
4, as a G-type star, corrected only for Galactic extinction and the
distance modulus, would be too faint, at ${M_V}^0 \approx -4.5$ mag.
We therefore consider the possibility that Star 4 is a yellow
hypergiant to be unlikely.

Star 4, however, also could be a reddened early-type supergiant.  We
have already pointed out the similarity of its spectrum to that of
$\eta$ Car, so it is plausible for us to assume that Star 4 could be a
LBV seen through appreciable extinction, as is the case for $\eta$
Car.  Even with the color offset from the stellar tracks, the star's
locus in Figure 5 does, in fact, appear consistent with a blue star
experiencing an extinction of $A_V \approx 4$ mag.  We note that it
is reasonable to assume that the reddening is variable across $\sim$18
pc at the SN site.  Unfortunately, we do not possess another observed
color (such as $U-B$) or stellar spectral features for these four stars
to help break some of the degeneracy between the stars' possible
intrinsic colors and the amount of reddening in the environment.

All four stars are somewhat redder in $B-V$ and/or bluer in $V-I$ than
we might expect for the inferred stellar types, possibly due to underestimation
of the uncertainties in the observed colors and to errors in the color transformation.
(For Stars 1 and 2 using, for example, SYNPHOT and the BPGS Atlas, we can transform the
flight system colors to $V-I=2.1$--2.3 and $B-V \gtrsim 1.7$ mag, putting the
stars closer to the supergiant track.)
Although we cannot readily account for the color offset for Star 3,
at least some of the offset for Star 4 ($\sim 0.2$ mag) from the
supergiant track likely results from contamination of the continuum light in the broad
F606W band by the significant emission at H$\alpha$.  We can explore this
further using SYNPHOT:
normalizing a 32,000~K blackbody (or, equivalently, a late O-type or early B-type
supergiant) spectrum to an unreddened F814W magnitude of 19.8
(absolute magnitude
${M_I}^0 \approx -7.7$
for a luminous supergiant at the distance of
NGC 2403), then applying a reddening
$E(B-V)=1.26$ mag ($A_V = 4$ mag),
and, finally, adding a synthetic
(Gaussian-shaped) H$\alpha$ emission line with the characteristics of
the observed emission line in Figure 2 and the observed F658N magnitude for
Star 4 in Table 2, results in a good fit to the observations, to within the uncertainties,
for the F475W, F606W, and F814W magnitudes of 24.13, 23.09, and 22.14, respectively.

Without the H$\alpha$ emission in the model, the F475W and F814W magnitudes, of
course, remain relatively unchanged; however, the F606W magnitude is then 23.24.
Following the transformations in Sirianni et al.~(2005), we find $B-V=1.00$ and
$V-I=1.34$ mag.  
(The $B-V$ and $V-I$ colors, 0.95 and 1.30 mag, respectively,
for the model derived using SYNPHOT are unaffected by the H$\alpha$ emission,
which is outside the $V$ passband.) These colors now place the locus 
of Star 4 much closer both to the supergiant track and to
being along the vector for a reddened luminous blue star.  The H$\alpha$ emission
clearly leads through the bandpass tranformations to a redder
$B-V$ and a bluer $V-I$ color (exactly the trend seen in Figure
5) than without the emission.

In light of these arguments,
we consider it quite
plausible that, like $\eta$ Car, Star 4 is a very luminous O-type or
early-B star hidden behind a dusty ($A_V \approx 4$ mag) shell or
LBV-like nebula.  We argue that this is the most likely scenario,
given the star's brightness, color, and the characteristics of the
H$\alpha$ line emission.  The shell, analogous to $\eta$ Car's
Homunculus, was possibly ejected during the 1954 super-outburst.  We
therefore assume the unreddened colors for the model star,
$(B-V)_0=-0.22$ and $(V-I)_0=-0.21$ mag, and the model
extinction-free absolute magnitude, ${M_V}^0 \approx -8.0$.
Including the measured uncertainties in the corresponding
flight-system F475W$-$F606W and F606W$-$F814W colors and F606W magnitude from
Table 2, we can place Star 4 in color-magnitude diagrams shown in
Figure 6, along with the possible properties of the other three stars
in the environment.

For comparison we show in Figure 6 stellar evolutionary tracks from
Lejeune \& Schaerer (2001), with enhanced mass loss, for a range of
stellar masses.  
At the position of SN 1954J/V12 in the galaxy
($105{\farcs}8$ from the nucleus, or galactocentric distance
$R\approx 1.6$ kpc) the metallicity could be somewhat subsolar:
From a study of the metallicity gradient in NGC 2403
by Garnett et al.~(1997), the value of log(O/H)+12 at this distance is 8.6 dex,
which is a factor 1.58 below solar metallicity
(the solar log(O/H)+12 abundance is 8.8 dex; Grevesse \& Sauval 1998).  
However, the first available 
subsolar model tracks from Lejeune \& Schaerer are for $Z=0.008$, which is 
a factor 2.5 below solar metallicity ($Z=0.02$), and we consider this too low 
to adequately represent the SN 1954J/V12 environment.  Therefore, we choose the
model tracks with solar metallicity (although we note below the ramifications
if we used the $Z=0.008$ tracks instead).
We conclude from Figure 6 that Stars 1 and 2 are
either yellow supergiants of $\sim 20\ M_{\odot}$ or, more likely,
lower-mass ($\sim 15\ M_{\odot}$, or $\lesssim 12\ M_{\odot}$ for the
$Z=0.008$ track) red supergiants, whereas Star 3
could be a blue main-sequence dwarf with a range of possible (likely,
low) masses.  More importantly, we conclude that Star 4 has a mass
$M \gtrsim 25\ M_{\odot}$ (or $M \gtrsim 20\ M_{\odot}$ for the $Z=0.008$ 
track; the mass is most constrained in Figure 6b).
Unfortunately, the photometric uncertainties do not allow
us to more rigorously constrain the mass of Star 4.  However, the
spectral and photometric similarities to $\eta$ Car, and its implied
high absolute $V$ magnitude, argue that the star's mass is likely
closer to the upper range of massive stellar tracks shown in Figure 6.
We therefore find it highly suggestive that this star is the very
luminous, massive survivor of the SN 1954J/V12 eruption.

What about the stellar environment of SN 1954J/V12, relative to
the environments of $\eta$ Car and other LBVs?  How likely is it that these
four stars constitute an association of massive stars spread over
$\sim 1{\farcs}6$ ($\sim 24$ pc)?  The star $\eta$ Car is
a member of the populous OB association Trumpler 16, part of a much larger
($\sim 50$ pc) site of intense massive star formation in the Carina Nebula
(Walborn 1995).  P Cygni appears to be a member of a rich young cluster
$\sim 18$ pc northeast of the cluster IC 4996 and part of the larger Cyg OB1
association (Turner et al.~2001).  SN 1961V also sits in an environment of luminous,
presumably massive supergiants over $\sim 300$ pc (Filippenko et al.~1995;
Van Dyk et al.~2002).
Although not an $\eta$ Car analog, SN 1987A arose from a high-mass blue supergiant
and occurred in a loose, young (age $\sim$12~Myr) cluster (Panagia et al.~2000);
its closest known companion is a coeval B2III star (Scuderi et
al.~1996).  However, the LBV AG Car does not
seem to be associated with any cluster of luminous, massive stars over
$\sim 34$ pc (Hoekzema, Lamers, \& van Genderen 1992).  Therefore,
although the tendency is for the LBVs and $\eta$~Car analogs to occur in
rich, massive clusters spread over many pc, a range of environments exists for very
massive stars.  It is possible, then, for the four stars in the SN 1954J/V12
environment to be a sparse cluster of roughly coeval massive stars (see Figure
6).

What about the dusty nebula?  The profile of the source in the F658N
image is unresolved ($\sim$1.5 WFC1 pixels FWHM), which implies a
radius $R({\rm H}\alpha) \lesssim 1.1$ pc for the nebula.  We can
further constrain the size of the nebula from the luminosity in the
H$\alpha$ emission line and recombination theory, following Wagner et
al.~(2004).  Assuming an average $\sim 7 \times 10^{-18}$ erg s$^{-1}$
cm$^{-2}$ \AA$^{-1}$ and equivalent width $\sim$240 \AA\ for the line,
and correcting for extinction using the relation $A_{{\rm
H}\alpha}=0.81A_V$ (Viallefond \& Goss 1986), we derive an estimate of
the H$\alpha$ luminosity for the nebula $L({\rm H}\alpha) \approx 3.3
\times 10^{37}$ erg s$^{-1}$ at the distance of NGC 2403.  Assuming
the same properties for this nebula as did Wagner et al.~(2004) for
the nebula around SN 2000ch, we find $R({\rm H}\alpha) \approx 0.1$
pc, comparable to (within a factor of two of) the size of $\eta$ Car's
Homunculus (Humphreys \& Davidson 1994).

\section{Conclusions}

Through imaging and spectroscopy obtained with the Keck 10-m telescopes
and imaging with the ACS on-board {\sl HST}, we have confirmed that
the precursor star of SN 1954J/V12 in NGC 2403 has survived a
super-outburst, analogous to the 1843 eruption of $\eta$ Car.  It is
not a SN, in the classical sense, but an ``impostor'' masquerading as
a SN.  Furthermore, the ACS images, in combination with the Keck
spectroscopy, have revealed this quite rare $\eta$ Car analog as a
highly luminous, massive star ($M_{\rm initial} \gtrsim
25\ M_{\odot}$), surrounded by a dusty, Homunculus-like
nebula.  (Based on this conclusion for SN 1954J, we also now consider
it even more likely that the $\eta$ Car-like star detected in the SN
1961V environment by Chu et al.~2004 is, in fact, the survivor of that
event.)  The fact that the SN 1954J/V12 super-outburst survivor has
changed little in brightness and color over the last eight years
implies that the star is now in a relatively quiescent state.  Its
behavior in the recent past suggests that monitoring of the object
should continue into the future, in order to understand more about
this important, but rare, phenomenon in the evolution of massive stars.

\acknowledgements

We are grateful to Soeren Larsen with assistance in the calibration of his
images. Doug Leonard and Adam Riess provided assistance during the Keck imaging
run.  We thank Roberta Humphreys for illuminating discussions.  This
publication makes use of data products from the Two Micron All Sky Survey,
which is a joint project of the University of Massachusetts and the
IPAC/California Institute of Technology, funded by NASA and NSF. We have also
utilized the NASA/IPAC Extragalactic Database (NED) which is operated by the
Jet Propulsion Laboratory, California Institute of Technology, under contract
with NASA.  The work of A.V.F.'s group at UC Berkeley is supported by NSF grant
AST-0307894, as well as by NASA grants AR-9953 and AR-10297 from the Space
Telescope Science Institute, which is operated by AURA, Inc., under NASA
contract NAS5-26555. Some of the data presented herein were obtained at the
W. M. Keck Observatory, which is operated as a scientific partnership among the
California Institute of Technology, the University of California, and NASA; the
Observatory was made possible by the generous financial support of the
W. M. Keck Foundation. A.V.F. is grateful for a Miller Research Professorship
at U.C. Berkeley, during which part of this work was completed.


\begin{deluxetable}{lcccccc}
\tablenum{1}
\tablewidth{6.5truein}
\tablecolumns{7}
\tablecaption{Photometry of SN 1954J/V12}
\tablehead{
\colhead{UT date} & \colhead{JD} & \colhead{$U$} & \colhead{$B$} & \colhead{$V$}
& \colhead{$R$} & \colhead{$I$} \\
\colhead{} & \colhead{} & \colhead{(mag)} & \colhead{(mag)} & \colhead{(mag)} & \colhead{(mag)} & \colhead{(mag)}}
\startdata
1996 Oct. 20 & 2450377.1 & \nodata & \nodata         & 21.31(30) & 21.06(09)  & 20.04(09) \\
1997 Oct. 13 & 2450734.7 & 21.67(06) & 22.21(05) & 21.74(05) & 20.98(05)  & 20.14(07)  \\
1999 Feb. 14--21\tablenotemark{a} & 2451227.0 & 22.5(3) & 22.7(2)  & 21.9(3)      & 21.1(2)      & 20.9(2) \\
2004 Aug. 17\tablenotemark{b} & 2453235.0 & \nodata & 22.60(23)   & 21.67(15) & \nodata      & 20.12(12) \\
\enddata
\tablenotetext{}{Note: uncertainties in hundredths of a magnitude are indicated
in parentheses (except for the 1999 Feb. data from Smith et al.~2001, where
uncertainties in tenths of a magnitude are given).}
\tablenotetext{a}{From Smith et al.~(2001).}
\tablenotetext{b}{Magnitude estimates are synthesized from the brightnesses of Stars 1, 3, and 4 in the SN
environment as seen in the {\sl HST\/} images (see Table 2). Star 2, likely the ``appendage,'' as seen
from the ground, is spatially distinct
from the other three stars, so its contribution to the integrated light is not included.}
\end{deluxetable}


\begin{deluxetable}{cccccccc}
\tablenum{2}
\tablewidth{7.0truein}
\tablecolumns{7}
\tablecaption{The SN 1954J/V12 Environment\tablenotemark{a}}
\tablehead{
\colhead{Star} & \colhead{F475W} & \colhead{F606W} & \colhead{F814W} & \colhead{F658N\tablenotemark{b}} & \colhead{$B$}
& \colhead{$V$} & \colhead{$I$} \\
\colhead{} & \colhead{(mag)} & \colhead{(mag)} & \colhead{mag} & \colhead{(mag)} & \colhead{(mag)} & \colhead{(mag)} & \colhead{(mag)}}
\startdata
1 & 23.99(33) & 22.25(12) & 20.38(08) & 22.10(15) & 24.74(33) & 22.72(12) & 20.40(08) \\
2 & 24.08(34) & 22.30(12) & 20.56(07) & 22.20(17) & 24.85(34) & 22.76(14) & 20.58(08) \\
3 & 23.01(14) & 22.81(17) & 22.82(44) & 23.49(42)\rlap{\tablenotemark{c}} & 23.07(14) & 22.83(17) & 22.81(44) \\
4 & 23.89(35) & 22.88(19) & 22.22(26) & 21.00(05) & 24.29(35) & 23.08(21) & 22.21(26) \\
\enddata
\tablenotetext{a}{The magnitudes listed in the table are on the VEGAmag system, unless otherwise noted.  Uncertainties of hundredths of a magnitude are indicated in parentheses.}
\tablenotetext{b}{The magnitudes in the F658N bandpass are on the STmag system.}
\tablenotetext{c}{This measurement is an aperture magnitude, since DAOPHOT rejected the source
as too faint during PSF fitting.}
\end{deluxetable}


\begin{figure}
\figurenum{1}
\caption{SN 1954J/V12, as identified by Smith et al.~(2001), is
indicated by the arrow in the ({\it a}) $R$-band and ({\it b})
$I$-band images obtained at the Keck-I 10-m telescope on 1996 October 20.
The fiducial star, which was also included within the slit during the
2002 Keck-II ESI spectroscopy (see text), is labelled as ``A'' in panel
({\it a}).  The less-than-optimal quality in the $I$ image is
noticeable in panel ({\it b}), but the stars are still well detected.
Note that the apparent morphology of SN 1954J/V12 changes with
wavelength.}
\end{figure}

\clearpage

\begin{figure}
\figurenum{2}
\plotone{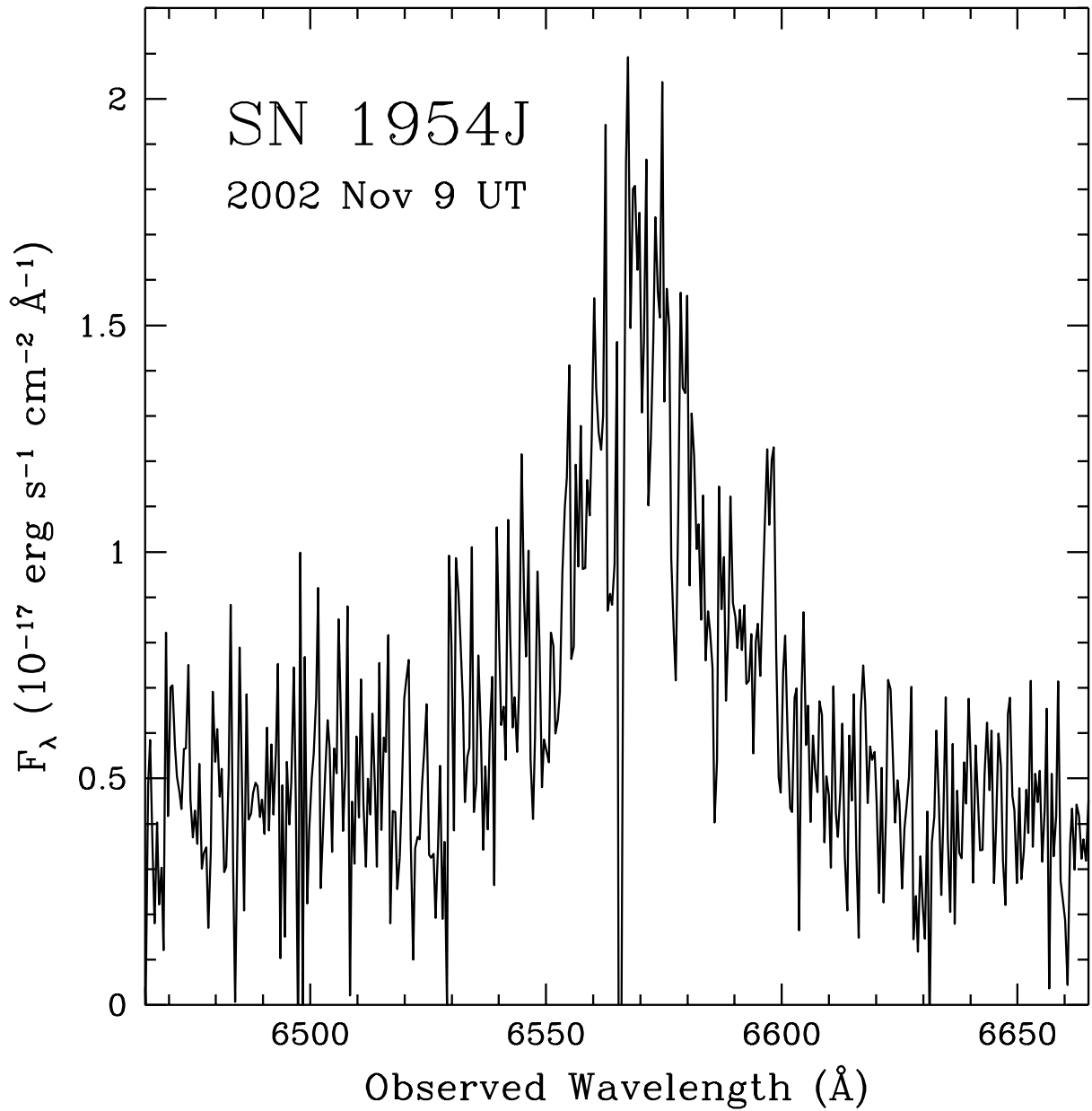}
\caption{The spectrum of the SN 1954J/V12 site obtained with ESI on
the Keck-II 10-m telescope on 2002 November 9.  The spectrum is noisy, but
clearly shows broad H$\alpha$ emission, with line width $v_{\rm exp}
\approx$ 700 km s$^{-1}$.}
\end{figure}


\begin{figure}
\figurenum{3}
\caption{({\it a}) The F606W, ({\it b}) F814W, and ({\it c}) F658N
images of the SN 1954J/V12 site obtained with the Advanced Camera for
Surveys/Wide Field Channel on-board {\sl HST\/} on 2004 August 17.  In
the inset to panel ({\it a}) we include a ``zoom'' of Figure 1a, showing
approximately the same field as shown in ({\it a}).  The four brightest stars in
the SN environment are labelled in the figure (see Table 2).}
\end{figure}

\clearpage

\begin{figure}
\figurenum{4}
\plotone{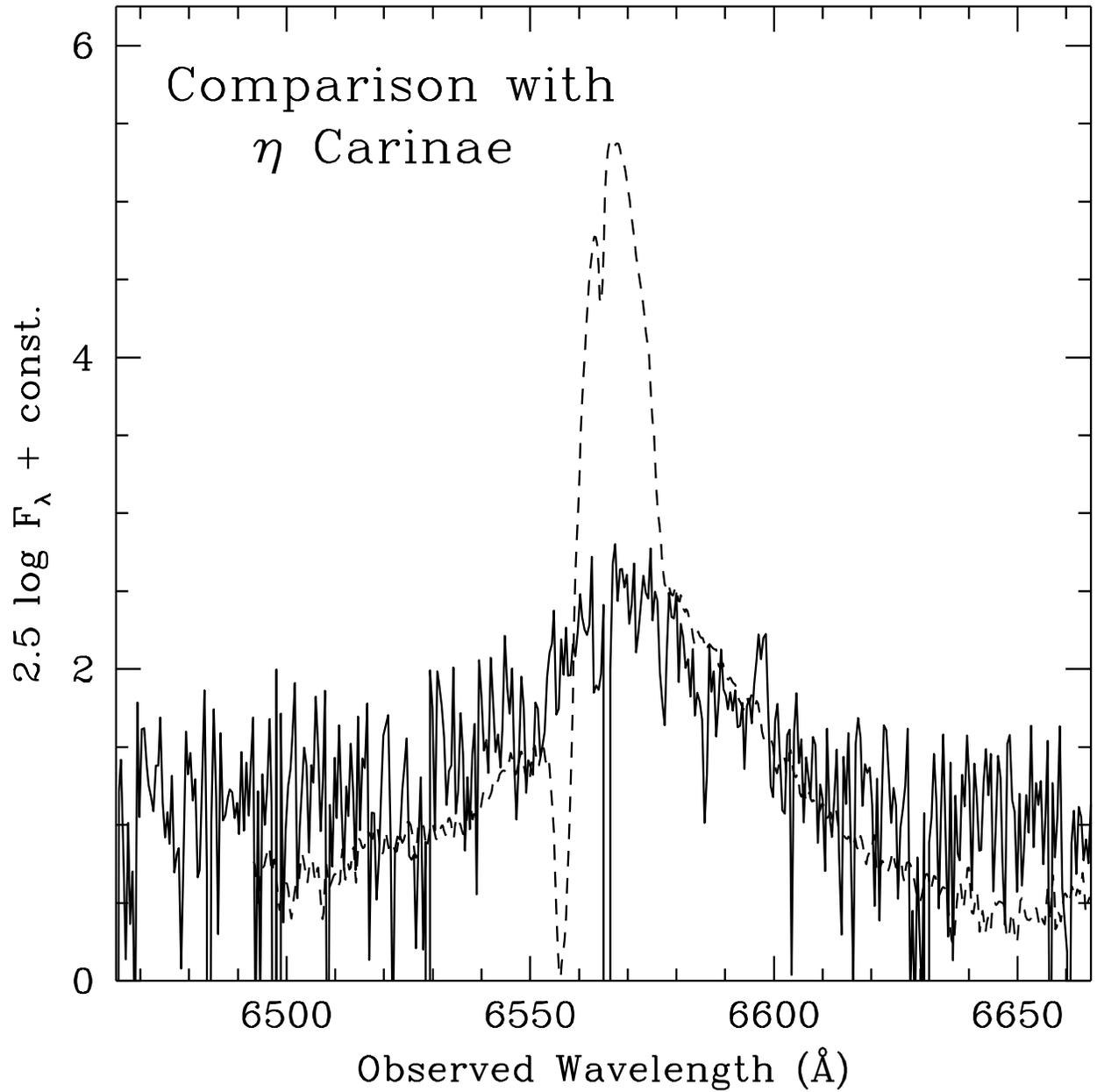}
\caption{A comparison of the Keck spectrum of SN 1954J/V12 ({\it solid line}),
shown in Figure 2, with a recent spectrum (1998 January 1) of $\eta$ Car
({\it dashed line}) obtained with STIS on-board {\sl HST\/} by program GO-9420
(PI: K.~Davidson).  The $\eta$ Car spectrum has been shifted in wavelength to
match the redshift of the SN 1954J/V12 host galaxy, NGC 2403.}
\end{figure}

\clearpage

\begin{figure}
\figurenum{5}
\plotone{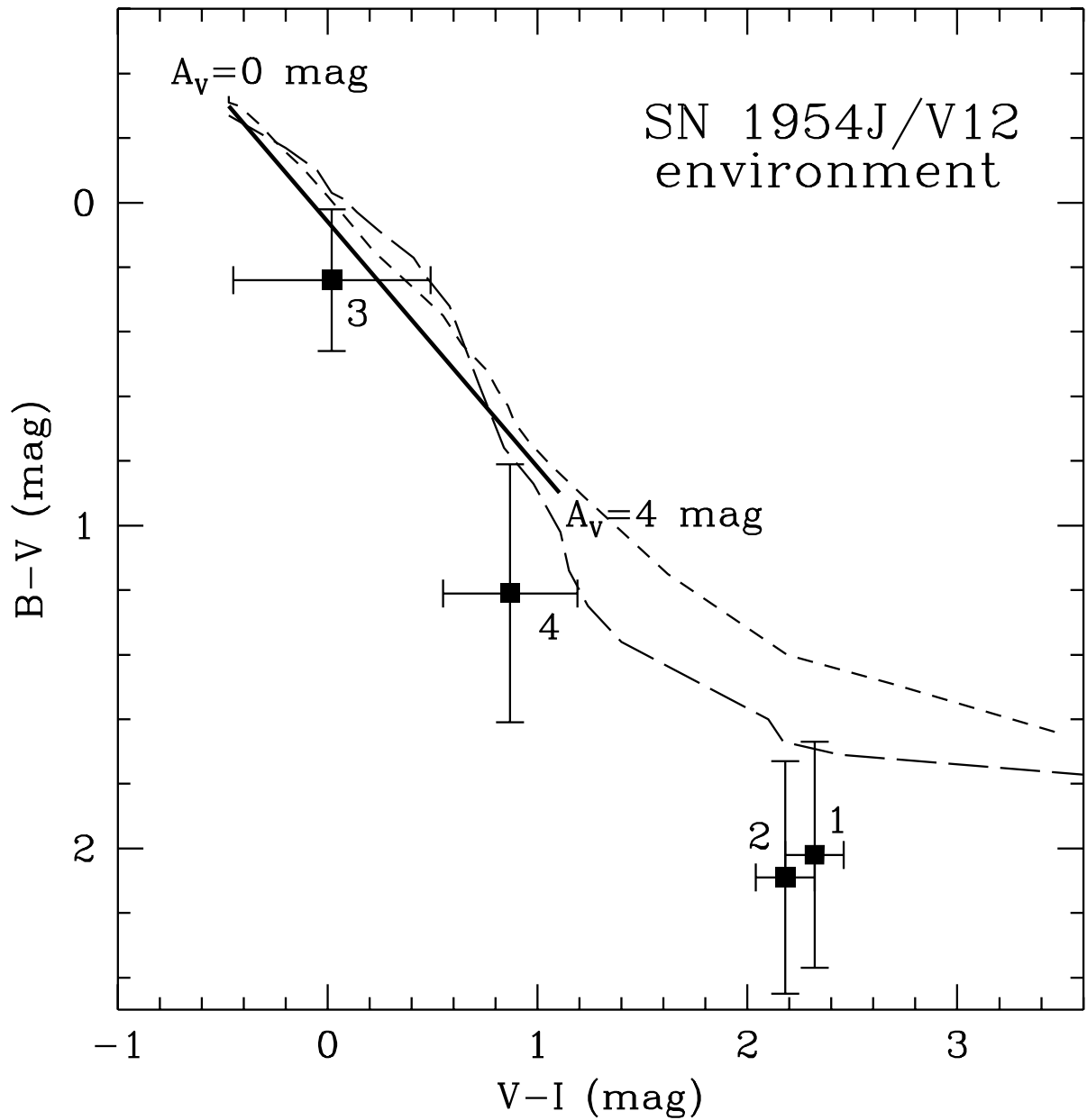}
\caption{The ($B-V$, $V-I$) color-color diagram for the four brightest
stars (labeled) at the SN 1954J/V12 site, based on the transformed
photometry in Table 2.  Also shown are the loci for supergiants ({\it
long-dashed line}) and zero-age main sequence (ZAMS) stars ({\it
short-dashed line}) from Drilling \& Landolt (2000), as well as the
reddening vector, from $A_V=0$ to $A_V=4$ mag, following a Cardelli et
al.~(1989) reddening law.}
\end{figure}

\clearpage

\begin{figure}
\figurenum{6}
\plotone{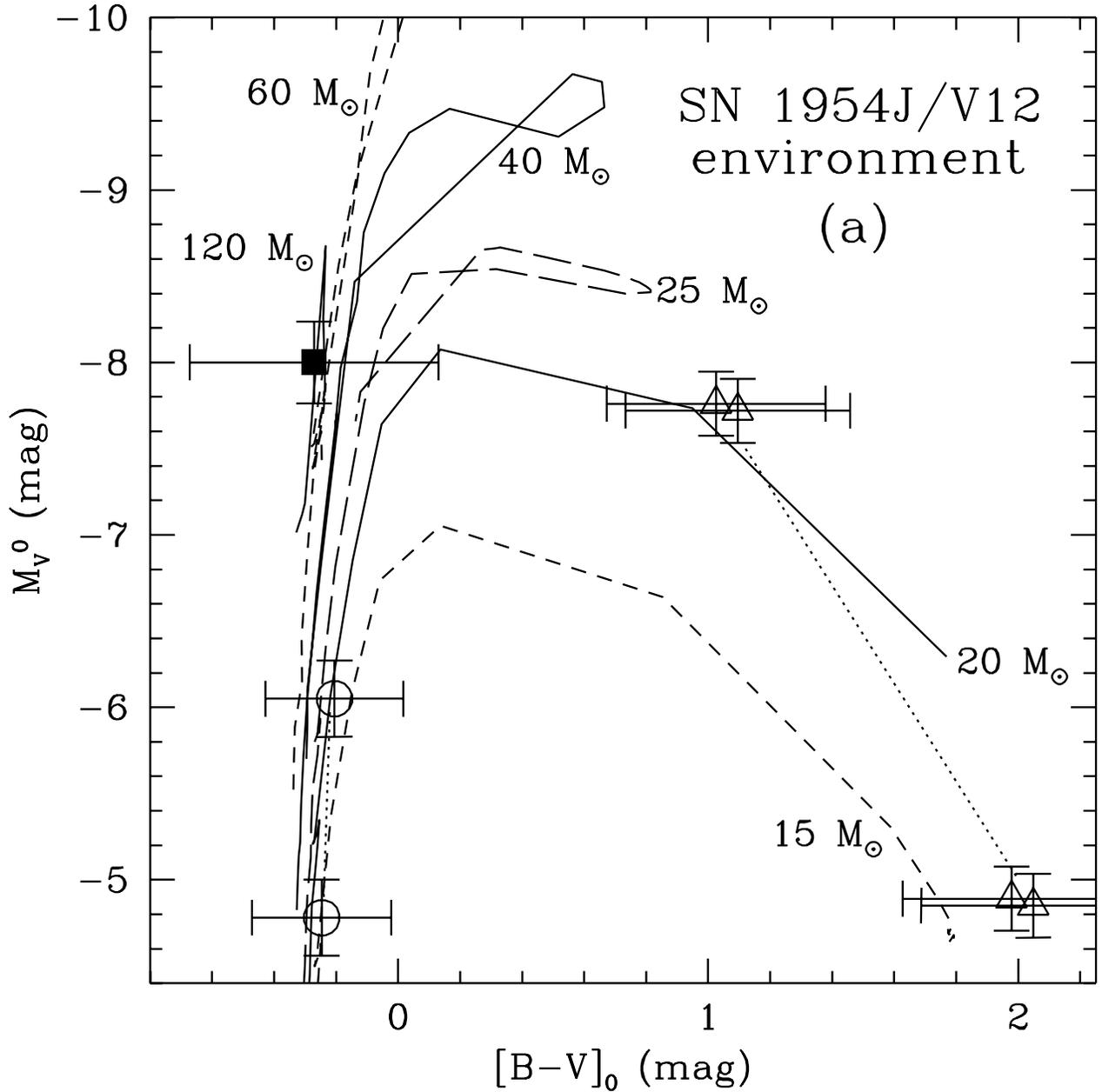}
\caption{Intrinsic color-magnitude diagrams, ({\it a}) (${M_V}^0$,
$[B-V]_0$) and ({\it b}) (${M_V}^0$, $[V-I]_0$), showing the loci of
the four stars in the SN 1954J/V12 environment (Stars 1 and 2,
{\it open triangles};
Star 3, {\it open circles}; and, Star 4, {\it filled square}).  
Also shown for comparison are the model stellar evolutionary
tracks from Lejeune \& Schaerer (2001), with enhanced mass loss and
solar metallicity ($Z=0.02$), for 120 $M_{\sun}$ ({\it solid
line}), 60 $M_{\sun}$ ({\it short dashed line}), 40 $M_{\sun}$ ({\it
solid line}), 25 $M_{\sun}$ ({\it long dashed line}), 20 $M_{\sun}$
({\it solid line}), and 15 $M_{\sun}$ ({\it short dashed line}).
{\it Dotted lines\/} connect the two
possible loci for Stars 1 and 2 and for Star 3.}
\end{figure}

\clearpage

\begin{figure}
\figurenum{6}
\plotone{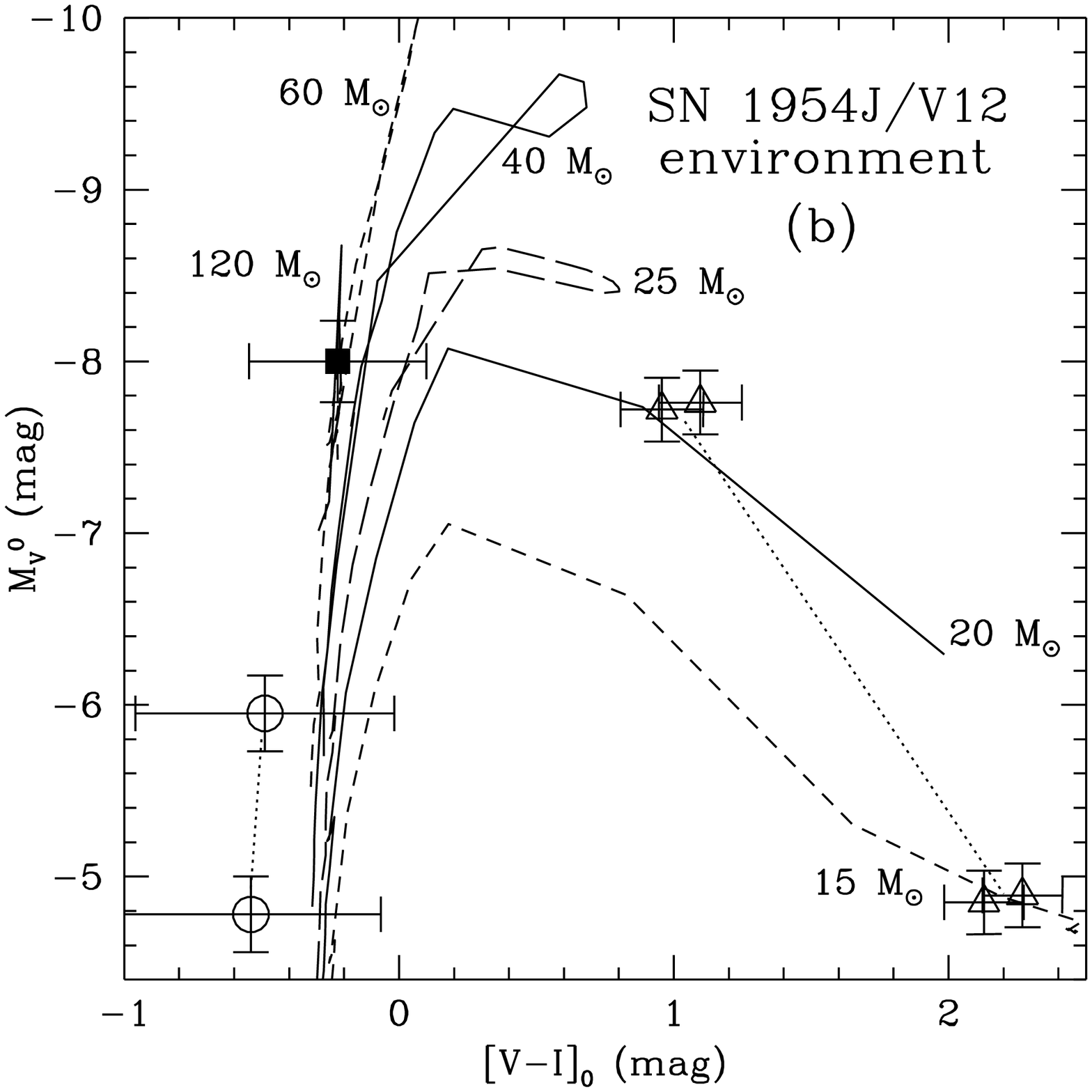}
\caption{(Continued.)}
\end{figure}

\end{document}